\documentclass[11pt]{article}
\usepackage{ifpdf}               
\usepackage{a4wide}             
\usepackage{amssymb}            
\usepackage{slashed}            
\ifpdf
\usepackage[pdftex]{graphicx}
\usepackage[pdftex,unicode,implicit]{hyperref}
\hypersetup{%
  pdftitle    = {IFT-UAM/CSIC-07-24: d=5 N=1 sugra vacua and quantum corrections; revised version},
  pdfkeywords = {supergravity, supersymmetric solutions, quantum corrections},
  pdfauthor   = {Patrick A.A. Meessen},
  pdfcreator  = {pdf\LaTeXe\ with package \flqq hyperref\frqq},
  pdfproducer = {pdf\LaTeXe\ with package \flqq hyperref\frqq},
  pdfpagemode = None,  
  pdffitwindow= true,  
  unicode     = true,
  plainpages  = false,
  colorlinks  = true,  
  citecolor   = black,
  urlcolor    = black,
  linkcolor   = black
}

\else
  \usepackage[dvips]{graphicx}
  \usepackage[unicode,implicit]{hyperref}

\fi
\begin{document}
\begin{flushright}
\small
IFT-UAM/CSIC-07-24\\
June 4, 2012
\normalsize
\end{flushright}
\begin{center}
{\large\bf All-order consistency of 5d sugra vacua}\\[.5cm]
{\bf Patrick~Meessen}\\[.2cm]
{\em Instituto de F\'{\i}sica Te\'orica UAM/CSIC, Universidad Aut\'onoma de Madrid\\
     M\'odulo C-XVI, Cantoblanco, E-28049 Madrid, Espa\~{n}a}\\[.5cm]
{\bf abstract}\\
\begin{quote}
{\small 
We show that the maximally supersymmetric vacua of minimal $d=5$ $N=1$ sugra remain
maximally supersymmetric solutions when taking into account higher order
corrections.
}
\end{quote}
\end{center}
\vspace{.5cm}
The question of whether a given supergravity solution is a consistent background
for string propagation to all orders in perturbation theory is an interesting
though hard one. It has of course long been known that pp-waves with flat
transverse space provide such backgrounds since all the scalar invariants vanish
identically -recently the class of exact sugra solutions with vanishing
scalar invariants was investigated in \cite{Coley:2007yx}- but they are exceptional.
\par
A class of sugra solutions for which a proof of all-order consistency is highly desireable are the
maximally supersymmetric solutions, and for most of them an answer is known:
in Ref. \cite{Kallosh:1998qs}, Kallosh {\&} Rajaraman by making use of superspace methods
showed the all-order consistency of $aDS_{2}\times S^{2}$ in minimal $N=2$ $d=4$ sugra, 
of $aDS_{5}\times S^{5}$ in type IIB, and also of $aDS_{4}\times S^{7}$ and $aDS_{7}\times S^{4}$
in M-theory. Since the associated Minkowski- and Kowalski-Glikmann solutions 
\cite{art:K_1,art:K_3,art:blau} are pp-waves, we must conclude that 
all the maximally supersymmetric solutions in minimal $N=2$ $d=4$, type IIB- or M-sugra
are all-order consistent.
\par
The way Kallosh {\&} Rajaraman attacked the problem leans heavily on the fact that 
the Riemann tensor and the field strengths are covariantly constant w.r.t. the Levi-Civit\`a
connection. This covariantly constancy, however, is due to the fact that the solutions
describe symmetric spacetimes, $G/H$, with $G$-invariant field strengths.
Interestingly, the vast majority of maximally supersymmetric solutions are described by symmetric
spaces, and one can envisage similar arguments to apply for their consistency.
\par
The strange ducks in the pond are the maximally supersymmetric solutions in $d=5$ sugra
\cite{Gibbons:1994vm,Kallosh:1996vy,Gauntlett:1998fz,Meessen:2001vx,Gauntlett:2002nw}.
Solutions such as the near-horizon-BMPV solution or the G\"odel
space, are {\em not} symmetric \cite{Alonso-Alberca:2002wr}: 
rather, they describe homogeneous, naturally reductive spacetimes with compatible fluxes. 
As such, there is a metric compatible connection that parallelises the Riemannan
tensor and the field strengths, but it is not the Levi-Civit\`a one. We then should
ask ourselves the question whether the maximally supersymmetric solutions of $d=5$ $N=1$ sugra are
all order exact or not,\footnote{
   Partial results have recently been obtained in Refs. \cite{Hanaki:2006pj,Castro:2007hc,Castro:2007sd} and
   these works instigated the current investigations: the symmetric solution $aDS_{2}\times S^{3}$ was shown 
   to be maximally supersymmetric in Ref. \cite{Castro:2007sd} and $aDS_{3}\times S^{2}$ in Ref. \cite{Castro:2007hc}
   in a theory with $\mathrm{Tr}(A\wedge^{2}R)$ corrections.
   This theory was constructed in \cite{Hanaki:2006pj} and they also derived the conditions for the existence of a 
   maximally supersymmetric $aDS_{5}$.
}
and how to attack the problem. The answer to the last question lies in the way one would construct,
and indeed constructs,
higher order sugra actions in lower dimensions, be they string inspired or not.
\par
Dealing with on-shell sypersymmetry in systems with higher orders is quite cumbersome: since
it is on-shell, the supersymmetry transformations `need to know about' the equations of motion,
so that {\em a priori} we would have to face the possibility of a lengthy Noether procedure
in order to construct the theory.\footnote{
  See {\em e.g.\/} \cite{Hyakutake:2007sm} and references therein, for the latest
  advances in the determination of the supersymmetric extention of the $R^{4}$ corrections in M-sugra. 
}
A possible off-shell formulation of supersymmetry evades this problem by fixing 
the supersymmetry transformations once and for all, so that the brunt of the effort
goes to the construction of the action. A similar problem occurs with the dependent fields
such as the spin connection. Indeed, trying to construct a higher order theory in the
first order prescription is cumbersome, as it calls for the elimination of the dependent
fields through the use of its equation of motion. In conclusion, we are looking for 
a supergravity description in which the dependent fields are already fixed and is 
as off-shell as possible.
\par
Having such a prescription, then, we can discuss the solutions to the equations defining 
unbroken supersymmetry, the so-called Killing spinor equation: 
as was to be expected, the solutions preserving all supersymmetries
are the ones from ordinary sugra. The more daunting task lies in showing that maximally
supersymmetric solutions always solve the equations of motion that can be derived from
an arbitrary sugralike action, in particular from an action that describes stringy corrections.
This can however be acomplished by making use of the Killing Spinor identities \cite{Kallosh:1993wx,Bellorin:2005hy}.
\par
In Sec.~(\ref{sec:SupConf}) we will apply the programme we sketched above to sugralike actions
constructed using the superconformal approach. In this approach, one starts out with locally
superconformal invariance, for which an off-shell formulation exists, to build superconformally
invariant actions. Once such an action is known, one imposes certain gauge choices in order to break
the superconformal invariance and to obtain (Poincar\'e) supergravities. Since the discussion
of the superconformal approach in Sec.~(\ref{sec:SupConf}) will be brief, but hopefully concise,
we refer the reader to Ref.~\cite{Bergshoeff:2004kh} as a possible starting point to the extensive
literature on the subject.
\par  
In order to declutter the technicalities, however, we will start by discussing and applying our strategy
in the less involved case of minimal $N=1$ $d=5$ sugra.
\section{Off-shell supersymmetry in Minimal $N=1$ $d=5$ sugra}
\label{sec:OffShell}
In Ref. \cite{Zucker:1999ej}, Zucker derived an off-shell formulation of minimal $d=5$ $N=1$ sugra
based on an off-shell multiplet of dimension $(48|48)$:\footnote{
  The minimal off-shell gravity multiplet has dimension $(40,40)$, but the quadratic sugra action
  based on this multiplet, leads to a constraint imposing the metric to be non-invertible.
  Further arguments for extention of the minimal multiplet are given in \cite{Howe:1981ev}.
}
it consists of the fields from the on-shell multiplet, namely the F\"unfbein $e_{\mu}{}^{a}$,
the vector field $A_{\mu}$ and the gravitino $\psi_{\mu}$, and a buch of auxiliar fields:
the fermions $\chi$ and $\lambda$ and the bosons $\varphi$, $V_{a}$, $v^{ab}$ and the 
$\mathfrak{su}(2)$-triplet fields $\vec{t}$ and $\vec{V}_{a}$.
The supersymmetry variations for purely bosonic configurations are
\begin{eqnarray}
  \label{eq:8a}
  \delta\Psi_{a} & =& \nabla_{a}\epsilon 
                \ +\ \textstyle{1\over 2\sqrt{3}}\left(
                         3\slashed{F}\gamma_{a} \ -\ \gamma_{a}\slashed{F}
                     \right)\epsilon 
                 \ +\ \textstyle{1\over 2}\gamma_{abc}\epsilon\ v^{bc}
                 \ -\ \textstyle{i\over 2}\vec{\sigma}\epsilon\ \vec{V}_{a}
                 \ +\ 2i\ \gamma_{a}\vec{\sigma}\epsilon\ \vec{t} \; , \\
    & & \nonumber \\
  \label{eq:8b}
  -2\delta\chi & =& \epsilon\ \varphi
               \ -\ 6i\ \vec{\sigma}\epsilon\ \vec{t} 
               \ -\ \gamma^{a}\epsilon\ V_{a}
               \ +\ \textstyle{i\over 2} \gamma^{a}\vec{\sigma}\epsilon\ \vec{V}_{a}
               \ -\ 2\slashed{v}\epsilon \; ,\\
     & & \nonumber \\
  4\delta\lambda & & \epsilon\ \left[
                         \nabla_{a}V^{a} \ +\ \textstyle{1\over 4}\vec{V}_{a}\cdot\vec{V}^{a}
                         \ -\ V_{a}V^{a} \ +\ \textstyle{5!\over 2}\vec{t}\cdot\vec{t}
                         \ +\ \varphi^{2} \ +\ 2v_{ab}v^{ab} \ +\ \textstyle{4\over \sqrt{3}}v^{ab}F_{ab}
                      \right] \nonumber \\
     & & -2\gamma_{a}\epsilon\ \left(
                 \nabla_{b}v^{ab} \ -\ \textstyle{1\over 2}\varepsilon^{abcde}v_{bc}v{de}
                 \ -\ \textstyle{2\over \sqrt{3}}\varepsilon^{abcde}F_{bc}v_{de} 
            \right) \nonumber \\
     & & +\gamma^{abc}\epsilon\ \nabla_{a}v_{bc} 
     \; +\; \gamma^{ab}\epsilon\ F_{a}{}^{c}v_{bc} \; . \label{eq:8c}
\end{eqnarray}
The analysis for the existence of maximally supersymmetric solutions of the above off-shell Killing spinor equation
is straightforward: from Eq. (\ref{eq:8b}), we see that 
$\varphi \ =\ \vec{t}\ =\ V_{a}\ =\ \vec{V}_{a}\ =\ v_{ab}\ =\ 0$ which automatically
trivializes Eq. (\ref{eq:8c}). 
Eq. (\ref{eq:8a}) then reduces to 
\begin{equation}
  \label{eq:9}
       0 \; =\; \nabla_{a}\epsilon 
                \ +\ \textstyle{1\over 2\sqrt{3}}\left(
                         3\slashed{F}\gamma_{a} \ -\ \gamma_{a}\slashed{F}
                     \right)\epsilon \; ,
\end{equation}
which is nothing but the Killing spinor equation for minimal on-shell $N=1$ $d=5$ sugra.
Clearly, the configurations solving the above equations are the ones enumerated by
Gauntlett {\em et al.} in Ref. \cite{Gauntlett:2002nw}, but we must ask ourselves
whether they automatically solve the equations of motion that can be derived from an
off-shell action based on Zucker's formulation.
\par
The way we want to show this to be the case was pioneered by Kallosh {\&} Ort\'{\i}n \cite{Kallosh:1993wx},
and first used to show that some e.o.m.s are implied by susy in Ref. \cite{Bellorin:2005hy}. This approach
goes by the name of {\em Killing Spinor Identities} and exploits the fact that 
the invariance of an action
under a (super)symmetry implies the identity (introducing a superset of fields $\Phi^{A}=\{ B^{a},F^{\alpha}\}$)
\begin{equation}
  \label{eq:3}
  \delta\mathcal{S} \; =\; \int_{5}\; \delta\Phi^{A}\ 
                 \frac{\delta\left( \sqrt{g}\mathcal{S}\right)}{\sqrt{g}\delta\Phi^{A}}
         \; =\; \int_{5}\ \delta\Phi^{A}\ \mathcal{E}_{A} \;\; \longrightarrow\;\;
  0 \; =\; \delta\Phi^{A}\, \mathcal{E}_{A}(\Phi ) \; ,
\end{equation}
where we introduced the notation in which the equation of motion for a field $\Phi^{A}$ is written
as $\mathcal{E}_{A}(\Phi )\ =\ 0$.
If we then consider the functional derivative of the last equation in (\ref{eq:3})
w.r.t. some fermion field and evaluate the resulting identity for purely bosonic configurations
that solve the Killing spinor equations, 
{\em i.e.\/} $F^{\alpha}\ =\ \delta_{\epsilon}F^{\alpha}\mid_{F=0}\ =\ 0$, we end up with
\begin{equation}
  \label{eq:8}
  0 \; =\; \left. \frac{\delta}{\delta F^{\beta}}\left[ \delta_{\epsilon}B^{a}\right]\right|_{F=0}\; \mathcal{E}_{a} \; .  
\end{equation}
This equation is the Killing Spinor Identity and must hold for any supersymmetric system.
\par
Let us start analysing the implications of the KSIs by calculating the one w.r.t. the auxiliar field $\lambda$. 
A short calculation results in
\begin{equation}
  \label{eq:11}
  0 \; =\; \mathcal{E}_{i}(\vec{t})\ \bar{\epsilon}\sigma^{i}
    \ -\ 2i\ \mathcal{E}(\varphi )\ \bar{\epsilon}
    \ -\ 2i\ \mathcal{E}^{a}(V)\ \bar{\epsilon}\gamma_{a}
    \ +\ 2i\ \mathcal{E}_{ab}(v)\ \bar{\epsilon}\gamma^{ab}
    \ -\ 4\ \mathcal{E}_{i}^{a}(\vec{V})\ \bar{\epsilon}\gamma_{a}\sigma^{i} \; .
\end{equation}
Since we are interested in maximally supersymmetric solutions, the above identity holds
for all $\epsilon$ which together with the properties of the $\gamma$- and $\sigma$-matrices
implies that
\begin{equation}
  \label{eq:16}
  0 \; =\; \mathcal{E}_{i}(\vec{t})\; =\; \mathcal{E}(\varphi )\;
       =\; \mathcal{E}^{a}(V)\; =\;  \mathcal{E}_{ab}(v)
    \; =\; \mathcal{E}_{i}^{a}(\vec{V})\; .
\end{equation}
In ordinary language this means that maximally supersymmetric solutions of the off-shell 
Killing spinor equations automatically solve the e.o.m. of the auxiliar fields.
The only non-trivial e.o.m.s
that remain are the ones for the bosonic on-shell fields, and they can be derived
form the gravitino KSI
\begin{equation}
  \label{eq:12}
  0 \; =\; 4\ \mathcal{E}_{a}^{\mu}(e)\ \bar{\epsilon}\gamma^{a} 
    \; +\; \sqrt{3}\ \mathcal{E}^{\mu}(A)\ \bar{\epsilon} \; ,
\end{equation}
where we already used the results in (\ref{eq:16}).
Applying the same reasoning as before, we reach the conclusion that we also
identically solve the `Einstein' and the `Maxwell' equations.
\par
The fact that the KSIs oblige the e.o.m.s to be identically satisfied for maximally supersymmetric
configurations should have been expected: indeed, since for maximally supersymmetric configurations
we can factor out the explicit appearance of the Killing spinor, $\epsilon$, in the KSIs and decompose
the latter into independent tensor-structure blocks. 
The crux of the matter is that since the fields are distinguishable due to their symmetry properties,
such as R-symmetry, there can only be one e.o.m. per block, whence a maximally supersymmetric configuration
always solves the e.o.m.s.
\par
The conclusion thus far is that if we use Zucker's off-shell multiplet to construct
effective actions, then the maximally supersymmetric solutions are the ones from 
$d=5$ $N=1$ sugra and whatever action action one writes down, they always solve
the corresponding equations of motion.
How does coupling minimal sugra to matter multiplets change this picture?
\section{Vacua and coupling to vector multiplets}
\label{sec:SupConf}
The field content of on-shell $d=5$ $N=1$ sugra coupled to $n$ vector multiplets\footnote{
  Hypermultiplets can be introduced as well, with the same result, but will not be treated
  in order to make the discussion a tad lighter. Also, in this section we shall follow 
  \cite{Hanaki:2006pj}'s conventions.
} 
is a F\"unfbein $e_{\mu}{}^{a}$, a (symplectic-Majorana) gravitino $\psi_{\mu}^{i}$ ($i=1,2$), 
$n+1$ vector fields $A_{\mu}^{I}$ ($I=1,\ldots ,n+1$) and $n$ scalars $\phi$: the scalars
parametrise a {\em very special manifold} through the $n+1$ sections $h^{I}(\phi )$ that are constrainted
to satisfy 
\begin{equation}
  \label{eq:17}
  \mathcal{C}_{IJK}\ h^{I}\ h^{J}\ h^{K} \; =\; 1 \; ,
\end{equation}
where $\mathcal{C}$ is a constant, completely symmetric 3-tensor.
\par
In order to arrive at this on-shell sugra by means of the superconformal approach one starts
by introducing one Weyl multiplet, $n+1$ vector multiplets and one hypermultiplet.
The Weyl multiplet is the superconformal analogue of the graviton-multiplet and consists of 
the F\"unfbein $e_{\mu}^{a}$, the gravitino $\psi_{\mu}^{i}$, the vectors $b_{\mu}$ and $\mathtt{V}_{\mu}^{(ij)}$, 
a symplectic-Majorana spinor $\xi$ and the scalars $\mathtt{T}_{ab}$ and $\mathtt{D}$.
The $n+1$ superconformal vector multiplets consist of vectors $A_{\mu}^{I}$, gaugini $\Omega^{Ii}$,
the {\em unconstrained} scalars $h^{I}$ and the auxiliar fields $\mathtt{Y}^{I(ij)}$. The hypermultiplet, then,
consists of scalars $\mathsf{A}^{i}_{j}$ and spinors $\zeta_{i}$ constrained by suitable reality conditions.
\par
The superconformal Killing spinor equations, {\em i.e.} the variation of the fermionic fields under
supersymmetry variations with parameter $\epsilon$ and conformal-supersymmetries with parameter $\eta$
evaluated for vanishing fermions, then read
\begin{eqnarray}
  \label{eq:5}
  \delta\psi_{a} & =& \mathcal{D}_{a}\epsilon \; +\ _{a}\slashed{\mathtt{T}}\epsilon 
    \; -\ \gamma_{a}\eta \; . \\
    & & \nonumber\\
  \label{eq:5b}
  \delta\chi & =& \mathtt{D}\epsilon \ -\ 2\gamma^{c}\gamma^{ab}\epsilon\ \mathcal{D}_{a}\mathtt{T}_{bc}
    \ -\ 2\gamma^{a}\epsilon\ \varepsilon_{abcde}\mathtt{T}^{bc}\mathtt{T}^{de}
    \ +\ 8\slashed{\mathtt{T}}\eta
    \ +\ 2\slashed{R}(\mathtt{V})\epsilon  \; , \\
    & & \nonumber \\
  \label{eq:5c}
  \delta\Omega^{I} & =& \mathtt{Y}^{I}\epsilon 
    \ -\ h^{I}\eta
    \ -\ \textstyle{1\over 2}\slashed{\mathcal{D}}h^{I}\ \epsilon
    \ -\ \textstyle{1\over 2}\slashed{F}^{I}\epsilon \; ,\\
    & & \nonumber \\
  \label{eq:5d}
  \delta\zeta^{i} & =& \slashed{\mathcal{D}}\mathsf{A}^{i}_{j}\ \epsilon^{j}
             \ -\ 2 \slashed{\mathtt{T}}\epsilon^{j}\ \mathsf{A}^{i}_{j}
             \ +\ 3\eta^{j}\ \mathsf{A}^{i}_{j} \; .
\end{eqnarray}
where the $\mathfrak{su}(2)$ indices of $\epsilon$, $\eta$, $\Omega^{I}$, $R(\mathtt{V})$ and $\mathtt{Y}^{I}$
are implict but present. In the above formulas we used
\begin{equation}
  \label{eq:13}
  \mathcal{D}\epsilon \; =\; \nabla\epsilon 
                      \ +\ \textstyle{1\over 2}\gamma\ \slashed{b}\epsilon 
                      \ +\ \mathtt{V}\epsilon \;\;\; ,\;\;\;
  \mathcal{D}\mathtt{T}_{ab} \; =\; \nabla\mathtt{T}_{ab} \ -\ b\ \mathtt{T}_{ab}\; ,
\end{equation}
\par
At this point we must discuss the gauge fixings in order to get rid off the fields
that are not part of the on-shell sugra. This process actually consists of two parts as we must not only
break the superconformal symmetry down to Poincar\'e symmetry, but also get rid of the auxiliar fields.
A clear exposition of the traditional gauge fixing program in sugra is given in Ref. \cite{Bergshoeff:2004kh},
which has the advantage of using physically sound criteria to select gauge fixings. One of these concerns
the normalisation of the Einstein-Hilbert term: the gauge fixing they impose is that the normalisation of the Einstein-Hilbert
term is canonical, and together with the $\mathtt{D}$-e.o.m. this implies Eq. (\ref{eq:17}). 
In the generic case, however, this criterion is rather cumbersome:
instead we will follow Ref. \cite{Castro:2007sd}, which has the advantage that the gauge-fixed Killing spinor
equations have a simple form.\footnote{
   The spirit of this programme is the same as the canonical one as displayed in Fig. 1 of Ref. \cite{Bergshoeff:2004kh}.
}
\par
The r\^{o}le of the gauge fixings is to break the superconformal symmetry down to Poincar\'e symmetry, 
which in the case at hand means breaking dilatations ($D$), conformal translations ($K_{a}$),
the R-symmetry ($\mathfrak{su}(2)$) as well as the special supersymmetries ($S$). 
The conformal translations are broken by the $K$-gauge $b_{\mu}=0$, and the rest of the 
symmetries are broken by imposing conditions on the compensating hypermultiplet:
R-symmetry is broken by the condition $d\mathsf{A}=0$, which is consistent
with the condition $\mathsf{A}^{2}=-2$ for breaking the dilatational symmetry. S-symmetry,
then, is broken by the condition $\zeta^{i}=0$.
\par
At this point the fields that do not match up with the on-shell sugra are $\mathtt{T}$, 
$\mathtt{V}$, $\mathtt{D}$ and $\chi$
in the Weyl multiplet, the $\mathtt{Y}^{I}$ and one of the $h^{I}$ and $\Omega^{I}$ from the vector multiplets.
In sugra, most of them are auxiliar fields and are therefore to be eliminated by the use of their equation of motion.
In fact we should be a bit more specific: in sugra the equation of motion for $\mathtt{D}$,
abbreviated as $\mathcal{E}(\mathtt{D})=0$, imposes the constraint in Eq. (\ref{eq:17}), since $\mathtt{D}$
appears linearly in the action and hence acts as a Lagrange multiplier.
Similarly, $\mathcal{E}^{ab}(\mathtt{T})=0$ results in $-2\mathtt{T}= C_{IJK}h^{I}h^{J}F^{K}$ and integration
over $\chi$ rids us of the unwanted gaugino.
Lastly, the e.o.m. for $\mathtt{V}$ would, ignoring fermionic contributions, identify $\mathtt{V}$ with the pull-back
of the $\mathfrak{su}(2)$ connection characterising the quaternionic-K\"ahler manifold spanned by the 
{\em non-compensating} hypermultiplets.
\par
One of the most important implications of the gauge fixings is that in order for them to not break the 
ordinary supersymmetries, any supersymmetry transformation must be acompagnied by a
compensating $S$-symmetry transformation.
Indeed, the supersymmetry condition $\delta\zeta^{i} = 0$ means that 
\begin{equation}
  \label{eq:14}
  \eta \; =\; \textstyle{2\over 3}\ \slashed{\mathtt{T}}\ \epsilon 
        \ -\ \textstyle{1\over 3}\ \slashed{\mathtt{V}}\ \epsilon\; .
\end{equation}
Plugging the above result into eq.~(\ref{eq:5c}) and imposing maximal-supersymmetry we find
\begin{equation}
  \label{eq:1err}
  \begin{array}{lclclcl}
    \mathtt{Y}^{I} & =& 0 & \hspace{.5cm},\hspace{.5cm}&
    F^{I} & =& -\textstyle{4\over 3}\ h^{I}\ \mathtt{T} \\
    & &  & & & & \\
    dh^{I} & =& 0 & ,& \mathtt{V} & =& 0 \; ;
  \end{array}
\end{equation}
in particular, the $\mathrm{SU}(2)$ gauge-connection $\mathtt{V}$ and the auxiliar field $\mathtt{Y}^{I}$ vanish identically.
\par
The analysis of the remaining off-shell Killing spinor equations is then straightforward and results in
\begin{equation}
  \label{eq:18}
  \begin{array}{lclclcl}
    \mathtt{D} & =& \textstyle{8\over 3}\ \mathtt{T}_{ab}\mathtt{T}^{ab}   &\hspace{.5cm},\hspace{.5cm}& 
    0 & =& \nabla_{\mu}\epsilon \ +\ _{\mu}\slashed{T}\epsilon 
      \ -\ \textstyle{2\over 3}\ \gamma_{\mu}\slashed{\mathtt{T}}\epsilon \\
     & &  & & & & \\
    d\mathtt{T} & =& 0 & ,& 
    \nabla_{b}\mathtt{T}^{ba} & =&  \textstyle{1\over 3}\varepsilon^{abcde}\mathtt{T}_{bc}\mathtt{T}_{de}\; .\\
   \end{array}
\end{equation}
The last three equations mean that the maximally supersymmetric solutions
are once again given by the ones from minimal on-shell sugra with $\mathtt{T}$ playing the r\^{o}le of
the graviphoton's field-strength.
\par
Let us start the discussion of the KSIs by calculating the one for the auxiliar spinor $\chi$, namely
\begin{eqnarray}
  \label{eq:19}
  0 & =& \mathcal{E}_{ab}(\mathtt{T})\, +\, 4\mathtt{T}_{ab}\ \mathcal{E}(\mathtt{D}) \; , \\
  \label{eq:19b} 
  0 & =& \nabla_{\mu}\mathcal{E}(\mathtt{D}) \; ,\\
  0 & =& h^{I}\mathcal{E}_{Iij}(\mathtt{Y}) \;\; =\;\;
         \mathcal{E}^{\mu}_{ij}(\mathtt{V}) \; .
\end{eqnarray}
This result may, seeing the similar result in the foregoing section, seem strange, yet it makes perfect sense: 
remember that in ordinary sugra $\mathcal{E}(\mathtt{D})=0$ imposes the constraint (\ref{eq:17}). The analysis
of the BPS equation, however, says nothing, and in fact cannot say anything, about the normalisation 
of the $h^{I}$. Rather, in order to embed the maximally supersymmetric vacua of minimal
sugra consistently, one has to solve $\mathcal{E}(\mathtt{D})=0$, after which
$\mathcal{E}(\mathtt{T})$ vanishes identically.
\par
The KSI w.r.t. the gaugini $\Omega^{I}$ then implies
\begin{equation}
  \label{eq:20}
  0\; =\; \nabla_{\mu}\mathcal{E}_{Iij}(\mathtt{Y})
   \; =\; \mathcal{E}_{I}(h)
   \; =\; \mathcal{E}^{\mu}_{I}(A) \; ,
\end{equation}
meaning that the e.o.m. for the gauge fields $A^{I}$ and the scalars $h^{I}$ are automatically satisfied.
The above KSI does, however, mean that we should check explicitly whether $\mathcal{E}_{Iij}(\mathtt{Y})$
really vanishes for our solutions. But its index-structure implies that in order to construct it
we must always use $\mathtt{Y}$ or $\mathtt{V}$ and since they vanish for the vacua, we must conclude
that $\mathcal{E}(\mathtt{Y})=0$ for the vacua.
\par
The one equation left to check is the Einstein equation, which can only reside in the gravitino KSI. In fact
if we impose that we already solved all the other equations of motion, we automatically find that 
maximal supersymmetry implies that $\mathcal{E}^{\mu}_{a}(e)=0$.
\par
In conclusion, we see that the question about maximally supersymmetric solutions and their all-order
consistency reduces to an embedding problem that determines the constant values of the scalars $h^{I}$
in terms of the parameters determining the maximally supersymmetric solutions. 
As advertised in Ref. \cite{Alishahiha:2007nn}, the embedding formula $\mathcal{E}(\mathtt{D})=0$
defines a deformation of very special geometry which should have a profound influence on for example the attractor
mechanism.
\section*{Acknowledgements}
It is a pleasure to thank J. Bellor\'{\i}n, M. H\"ubscher, K. Landsteiner,
S. Vaul\`a and especially T. Ort\'{\i}n for fruitful discussions.
The author would furthermore like to thank P. Sloane for discussions which led to 
the discovery of a serious flaw in the original argumentation used in sec.~(\ref{sec:SupConf}).
This work has been supported in part by 
the {\em Fondo Social Europeo} through an I3P scholarship,
the Spanish Ministry of Science and Education grant FPA2006-00783, 
the Comunidad de Madrid grant HEPHACOS P-ESP-00346 
and by the EU Research Training Network \textit{Constituents,
Fundamental Forces and Symmetries of the Universe} MRTN-CT-2004-005104.

\end{document}